\documentclass{llncs}
\usepackage{hyperref}
\usepackage{version}
\usepackage{bdldb}

\newcommand{\doi}[1]{{doi:\href{http://doi.org/#1}{\nolinkurl{#1}}}\rmFullStop}

\newcommand{\eprintlnk}[1]{{\href{#1}{Electronic version}}\rmFullStop}
\renewcommand{\url}[1]{{\href{#1}{\nolinkurl{#1}}}\rmFullStop}
\newcommand*{\rmFullStop}{}

\begin{document}

\title{Belnap-Dunn Logic and Query Answering \\ 
       in Inconsistent Databases with Null Values}
\author{C.~A. Middelburg \\
        {\small ORCID: \url{https://orcid.org/0000-0002-8725-0197}}}
\institute{Informatics Institute, Faculty of Science, University of
           Amsterdam, \\
           Science Park~900, 1098~XH Amsterdam, the Netherlands \\
           \email{C.A.Middelburg@uva.nl}}

\maketitle

\begin{abstract}
This paper concerns an expansion of first-order Belnap-Dunn logic, 
called \pfoBDif, and an application of this logic in the area of 
relational database theory.
The notion of a relational database, the notion of a query applicable to 
a relational database, and several notions of an answer to a query with 
respect to a relational database are considered from the perspective of 
this logic, taking into account that a database may be an inconsistent
database and/or a database with null values. 
The chosen perspective enables among other things the definition 
of a notion of a consistent answer to a query with respect to a possibly 
inconsistent database without resort to database repairs.
For each of the notions of an answer considered, being an answer to a 
query with respect to a database of the kind considered is decidable.
\begin{classcode}
F.4.1, H.2.3, H.2.4
\end{classcode}
\end{abstract}

\section{Introduction}
\label{INTRO}

In the area of relational database theory, it is quite common since the 
1980s to take the view that a database is a set of formulas of 
first-order classical logic, a query is a formula of first-order 
classical logic, and query answering amounts to proving that a formula 
is a logical consequence of a set of formulas in first-order classical 
logic (see e.g.~\cite{Rei84a,GMN84a}).
The use of first-order classical logic makes it difficult to take into 
account the possibility that a database is an inconsistent database and 
the possibility that a database is a database with null values. 

In the logical view of relational databases, an inconsistent relational 
database is an inconsistent set of formulas in the sense that there is 
at least one conjunction of two formula of which one is the negation of 
the other that is a logical consequence of the set of formulas.
Because in classical logic every formula is a logical consequence of 
each such a conjunction, it becomes difficult to take the possibility 
that a database is inconsistent properly into account.
A logic in which not every formula is a logical consequence of each 
conjunction of two formula of which one is the negation of 
the other is called a paraconsistent logic.

The view on null values in relational databases taken in this paper can 
be described as follows:
(a)~in relational databases with null values, a single dummy value, 
called the null value, is used for values that are indeterminate,
(b)~a value that is indeterminate is a value that is either unknown or 
nonexistent, and
(c)~independent of whether it is unknown or nonexistent, no meaningful 
answer can be given to the question whether the null value and whatever 
value, including the null value itself, are the same.
Several variations on this view on null values in relational databases 
have been studied (see e.g.~\cite{Cod79a,Vas79a,IL84a}).
However, those variations are less basic and most of them give rise to 
similar problems in a logical view of relational databases with null 
values.

In the logical view of relational databases, a relational database with 
null values is an incomplete set of formulas in the sense that there
is at least one disjunction of two formula of which one is the negation 
of the other that is not a logical consequence of the set of formulas.
Because in classical logic each such a disjunction is a logical 
consequence of every set of formulas, it becomes difficult to take the 
possibility that null values occur in a database properly into account.
A logic in which not each disjunction of two formula of which one is the 
negation of the other is a logical consequence of every set of formulas 
is called a paracomplete logic.

In~\cite{Mid22b}, the notion of a relational database, the notion of a 
query applicable to a relational database, and several notions of an 
answer to a query with respect to a relational database are considered
from the perspective of \foLPif, an expansion of the first-order version 
of Priest's logic of paradox known as LPQ~\cite{Pri79a}. 
The possibility that a database is an inconsistent database is taken 
into account, but the possibility that a database is a database with 
null values is not taken into account. 
The reason for this is that \foLPif\ is paraconsistent, but not 
paracomplete.
There are many other paraconsistent logics.
The choice of \foLPif\ stems from the fact that its connectives and 
quantifiers are all familiar from classical logic and its logical 
consequence relation is very closely connected to the one of classical 
logic.

In~\cite{Mid23a}, \foBDif, an expansion of first-order Belnap-Dunn 
logic~\cite{AB63a} with classical connectives, is introduced and studied 
with the emphasis on 
(a)~the connection between the logical consequence relations of \foBDif\ 
and the version of classical logic with the same connectives and 
quantifiers and 
(b)~the definability in \foBDif\ of interesting non-classical 
connectives added to Belnap-Dunn logic in its expansions that have been 
studied earlier.
\foBDif\ is both paraconsistent and paracomplete.
Like for \foLPif, it holds for \foBDif\ that its connectives and 
quantifiers are all familiar from classical logic and its logical 
consequence relation is very closely connected to the one of classical 
logic.

An appendix of~\cite{Mid23a} goes briefly into an minor variation of 
\foBDif\ that covers terms with an indeterminate value.
This variation, called \pfoBDif, is also paraconsistent and 
paracomplete.
Moreover, its treatment of equations corresponds to the above-mentioned 
thought that no meaningful answer can be given to the question whether 
the null value and whatever value, including the null value itself, are 
the same value.  
All this means that both the possibility that a database is an 
inconsistent database and the possibility that a database is a database 
with null values can be properly taken into account when the notions 
considered in~\cite{Mid22b} are considered from the perspective of 
\pfoBDif. 

This is taken up in the current paper. 
In order to make this paper self-contained, the language and logical 
consequence relation of \foBDif\ as well as a sequent calculus proof 
system for \foBDif\ are presented.
\pfoBDif\ is introduced as a minor variation of \foBDif.
The notion of a relational database, the notion of a query applicable to 
a relational database, and several notions of an answer to a query with 
respect to a relational database are defined in the setting of \pfoBDif,
taking into account that a database may be an inconsistent database 
and/or a database with null values.
Like in~\cite{Mid22b}, the definitions concerned are based on those 
given in~\cite{Rei84a}.
Two notions of a consistent answer to a query with respect to a possibly
inconsistent relational database are introduced.
Like in~\cite{Mid22b}, one of them is reminiscent of the notion of a 
consistent answer from~\cite{Bry97a} and the other is reminiscent of the 
notion of a consistent answer from~\cite{ABC99a}.
 
The structure of this paper is as follows.
In Sections~\ref{LANGUAGE}, \ref{INTERPRETATION}, 
and~\ref{PROOF-RULES}, the language of \foBDif, the logical consequence 
relation of \foBDif, and a sequent calculus proof system for \foBDif\ 
are presented.
In Section~\ref{INDETERMINATE}, \pfoBDif\ is introduced as a minor 
variation of \foBDif.
In Sections~\ref{DATABASE} and~\ref{QUERY-ANSWERING}, relational 
databases and query answering with respect to a database that may be an
inconsistent database and/or a database with null values are considered 
from the perspective of \foBDif.
In Section~\ref{REMARKS}, some variations of the view on null values in
relational databases taken in this paper are briefly discussed.
In Section~\ref{CONCLUSIONS}, some concluding remarks are made.
Parts of the sections in which \foBDif\ is introduced overlap with parts 
of~\cite{Mid23a}.

\section{The Language of \foBDif}
\label{LANGUAGE}

In this section the language of the logic \foBDif\ is described.
First the notions of a signature and an alphabet are introduced and then 
the terms and formulas of \foBDif\ are defined for a fixed but arbitrary 
signature. 
Moreover, some relevant notational conventions are presented and some 
remarks about free variables and substitution are made.
In coming sections, the logical consequence relation of \foBDif\ and
a proof system for \foBDif\ are presented for a fixed but arbitrary 
signature.

\subsubsection*{Signatures and alphabets}

It is assumed that the following has been given:
(a)~a countably infinite set $\SVar$ of \emph{variables},
(b)~for each $n \in \Nat$, a countably infinite set $\Func{n}$ of 
\emph{function symbols of arity $n$}, and,
(c)~for each $n \in \Nat$, a countably infinite set $\Pred{n}$ of 
\emph{predicate symbols of arity $n$}.
It is also assumed that all these sets are mutually disjoint and 
disjoint from the set 
$\set{{=},\Not,\CAnd,\COr,\IImpl,\forall,\exists}$.

The symbols from 
$\Union \set{\Func{n} \where n \in \Nat} \union
 \Union \set{\Pred{n} \where n \in \Nat}$ 
are known as \linebreak[2] \emph{non-logical symbols}.
Function symbols of arity $0$ are also known as \emph{constant symbols} 
and predicate symbols of arity $0$ are also known as \emph{proposition 
symbols}.

A \emph{signature} $\vSigma$ is a subset of 
$\Union \set{\Func{n} \where n \in \Nat} \union
 \Union \set{\Pred{n} \where n \in \Nat}$. 
We write $\Func{n}(\vSigma)$ and $\Pred{n}(\vSigma)$, 
where $\vSigma$ is a signature and $n \in \Nat$, 
for the sets $\vSigma \inter \Func{n}$ and $\vSigma \inter \Pred{n}$, 
respectively. 

The language of \foBDif\ will be defined for a fixed but arbitrary 
signature~$\vSigma$.
This language will be called the language of \foBDif\ over $\vSigma$ or 
shortly the language of $\foBDif(\vSigma)$.
The corresponding logical consequence relation will be called the 
logical consequence relation of $\foBDif(\vSigma)$ and a corresponding 
proof system will be called a proof system for $\foBDif(\vSigma)$.

The \emph{alphabet} of the language of $\foBDif(\vSigma)$ consists of
the following symbols:
\begin{itemize}
\item
the variables from $\SVar$\,;
\item
the non-logical symbols from $\vSigma$\,;
\item
the \emph{logical symbols} of \foBDif, to wit:
\begin{itemize}
\item
the \emph{equality symbol} $=$\,;
\item
the \emph{falsity connective} $\False$\,;
\item
the \emph{negation connective} $\Not$\,;
\item
the \emph{conjunction connective} $\CAnd$\,;
\item
the \emph{disjunction connective} $\COr$\,;
\item
the \emph{implication connective} $\IImpl$\,;
\item
the \emph{universal quantifier} $\forall$\,;
\item
the \emph{existential quantifier} $\exists$\,.
\end{itemize}
\end{itemize}

\subsubsection*{Terms and formulas}

The language of $\foBDif(\vSigma)$ consists of terms and formulas.
They are constructed from the symbols in the alphabet of the language 
of $\foBDif(\vSigma)$ according to the formation rules given below.

The set of all \emph{terms of $\foBDif(\vSigma)$}, 
written $\STerm{\vSigma}$, is inductively defined by the following 
formation rules:
\begin{enumerate}
\item
if $x \in \SVar$, then $x \in \STerm{\vSigma}$;
\item
if $c \in \Func{0}(\vSigma)$, then $c \in \STerm{\vSigma}$;
\item
if $f \in \Func{n+1}(\vSigma)$ and 
$t_1,\ldots,t_{n+1} \in \STerm{\vSigma}$, then 
$f(t\sb1,\ldots,t_{n+1}) \in \STerm{\vSigma}$.
\end{enumerate}
The set of all \emph{closed terms of $\foBDif(\vSigma)$} is the subset of 
$\STerm{\vSigma}$ that can be formed by applying formation rules 2 and 3 
only.

The set of all \emph{formulas of $\foBDif(\vSigma)$}, 
written $\SForm{\vSigma}$, is inductively defined by the following 
formation rules:
\begin{enumerate}
\item
if $p \in \Pred{0}(\vSigma)$, then 
$p \in \SForm{\vSigma}$;
\item
if $P \in \Pred{n+1}(\vSigma)$ and 
$t_1,\ldots,t_{n+1} \in \STerm{\vSigma}$, then 
$P(t_1,\ldots,t_{n+1}) \in \SForm{\vSigma}$;
\item
if $t_1,t_2 \in \STerm{\vSigma}$, then $t_1 = t_2 \in \SForm{\vSigma}$;
\item
$\False \in \SForm{\vSigma}$;
\item
if $A \in \SForm{\vSigma}$, then $\Not A \in \SForm{\vSigma}$;
\item
if $A_1,A_2 \in \SForm{\vSigma}$, then 
$A_1 \CAnd A_2,\, A_1 \COr A_2,\, A_1 \IImpl A_2 \in \SForm{\vSigma}$;
\item
if $x \in \SVar$ and $A \in \SForm{\vSigma}$, then 
$\CForall{x}{A},\, \CExists{x}{A} \in \SForm{\vSigma}$.
\end{enumerate}
The set $\SAForm{\vSigma}$ of all 
\emph{atomic formulas of $\foBDif(\vSigma)$} is the subset of 
$\SForm{\vSigma}$ can be formed by applying formation rules 1--3 only.
The set $\SLit{\vSigma}$ of all \emph{literals of $\foBDif(\vSigma)$} is 
the subset of $\SForm{\vSigma}$ can be formed by applying formation 
rules 1--3 and 5 only.

We write $e_1 \equiv e_2$, where $e_1$ and $e_2$ are terms from 
$\STerm{\vSigma}$ or formulas from $\SForm{\vSigma}$, to indicate that
$e_1$ is syntactically equal to $e_2$.

In the coming sections, we will write \FOCL\ for the version of 
classical logic that has the same language as \foBDif.

\subsubsection*{Notational conventions}

The following will sometimes be used without mentioning (with or without 
decoration):
$x$~as a meta-variable ranging over all variables from 
$\SVar$,
$t$~as a meta-variable ranging over all terms from 
$\STerm{\vSigma}$, 
$A$ as a meta-variable ranging over all formulas from 
$\SForm{\vSigma}$, and
$\vGamma$ as a meta-variable ranging over all sets of formulas from
$\SForm{\vSigma}$.

The string representation of terms and formulas suggested by the 
formation rules given above can lead to syntactic ambiguities. 
Parentheses are used to avoid  such ambiguities.
The need to use parentheses is reduced by ranking the precedence of the 
logical connectives $\Not$, $\CAnd$, $\COr$, $\IImpl$.
The enumeration presents this order from the highest precedence to the
lowest precedence.
Moreover, the scope of the quantifiers extends as far as possible to
the right and
$\CForall{x_1}{\cdots \CForall{x_n}{A}}$ and
$\CExists{x_1}{\cdots \CExists{x_n}{A}}$ are usually written 
as $\CForall{x_1,\ldots,x_n}{A}$ and $\CExists{x_1,\ldots,x_n}{A}$,
respectively.

\subsubsection*{Free variables and substitution}

Free variables of a term or formula and substitution for variables in a 
term or formula are defined in the usual way.

Let $x$ be a variable from $\SVar$, $t$ be a term from 
$\STerm{\vSigma}$, and $e$ be a term from $\STerm{\vSigma}$ or a formula 
from $\SForm{\vSigma}$.
Then we write $\subst{x \assign t} e$ for the result of substituting the 
term $t$ for the free occurrences of the variable $x$ in~$e$, 
avoiding (by means of renaming of bound variables) free variables 
becoming bound in $t$.

\section{Truth and Logical Consequence in $\foBDif(\vSigma)$}
\label{INTERPRETATION}

In this section, the truth value of formulas of $\foBDif(\vSigma)$ 
and the logical consequence relation on sets of formulas of 
$\foBDif(\vSigma)$ will be defined.
This will be done using the logical matrix of $\foBDif(\vSigma)$.

First, the logical matrix of $\foBDif(\vSigma)$ is defined.
Next, the truth value of formulas of $\foBDif(\vSigma)$ and the logical 
consequence relation on sets of formulas of $\foBDif(\vSigma)$ are 
defined.
The truth value of formulas is defined with respect to 
(a)~a structure consisting of a set of values and an interpretation of 
each non-logical symbol and the equality symbol and
(b)~an assignment of values from that set of values to the variables.
Structures and assignments are introduced before the definition in 
question.

\subsubsection*{Matrix}

The interpretation of the logical symbols of $\foBDif(\vSigma)$, with the 
exception of the equality symbol, is given by means of a logical matrix.

In the definition of this matrix, $\VTrue$ (\emph{true}), 
$\VFalse$ (\emph{false}), $\VBoth$ (\emph{both true and false}), and 
$\VNeither$ (\emph{neither true nor false}) are taken as truth values.
Moreover, use is made of the partial order $\leq$ on the set 
$\set{\VTrue,\VFalse,\VBoth,\VNeither}$ in which $\VFalse$ is the least 
element, $\VTrue$ is the greatest element, and $\VBoth$ and $\VNeither$ 
are incomparable.
We write $\inf V$ and $\sup V$, 
where $V \subseteq \set{\VTrue,\VFalse,\VBoth,\VNeither}$, for the
greatest lower bound and least upper bound, respectively, of $V$ with 
respect to~$\leq$.

The \emph{matrix} of $\foBDif(\vSigma)$ is the triple 
$\langle \TValue, \DValue, \TFunct \rangle$, where:
\begin{itemize}
\item
$\TValue = \set{\VTrue,\VFalse,\VBoth,\VNeither}$;
\item
$\DValue = \set{\VTrue,\VBoth}$;
\item
$\TFunct$ consists of the following constant and functions: 
\[
\begin{array}{l@{\;\;}c@{\;\;}rcl}
\widetilde{\False} \in \TValue & 
\mathrm{defined\, by} &
\widetilde{\False} & = &
 \begin{array}[t]{l}
 \VFalse \;,
 \end{array}
\\[.5ex]
\widetilde{\Not} \mathbin{:} \TValue \to \TValue & 
\mathrm{defined\, by} &
\widetilde{\Not}(a) & = &
 \left \{
 \begin{array}{l@{\;\;}l}
 \VTrue  & \mathrm{if}\; a = \VFalse \\
 \VFalse & \mathrm{if}\; a = \VTrue \\
 a       & \mathrm{otherwise}\;,
 \end{array}
 \right.
\vspace*{.5ex} \\
\widetilde{\CAnd} \mathbin{:} \TValue \times \TValue \to \TValue & 
\mathrm{defined\, by} &
\widetilde{\CAnd}(a_1,a_2) & = &
 \begin{array}[t]{l}
 \inf \set{a_1,a_2}\;,
 \end{array}
\vspace*{.5ex} \\
\widetilde{\COr} \mathbin{:} \TValue \times \TValue \to \TValue & 
\mathrm{defined\, by} &
\widetilde{\COr}(a_1,a_2) & = &
 \begin{array}[t]{l}
 \sup \set{a_1,a_2}\;,
 \end{array}
\end{array}
\]
\[
\begin{array}{l@{\;\;}c@{\;\;}rcl}
\widetilde{\IImpl} \mathbin{:} \TValue \times \TValue \to \TValue & 
\mathrm{defined\, by} &
\widetilde{\IImpl}(a_1,a_2) & = &
 \left \{
 \begin{array}{l@{\;\;}l}
 \VTrue  & \mathrm{if}\; a_1 \notin \set{\VTrue,\VBoth} \\
 a_2 & \mathrm{otherwise}\;,
 \end{array}
 \right.
\vspace*{.5ex} \\
\widetilde{\forall} \mathbin{:}
 \mathcal{P}(\TValue) \diff \set{\emptyset} \to \TValue & 
\mathrm{defined\, by} &
\widetilde{\forall}(V) & = & \inf V\;,
\vspace*{.5ex} \\
\widetilde{\exists} \mathbin{:}
 \mathcal{P}(\TValue) \diff \set{\emptyset} \to \TValue & 
\mathrm{defined\, by} &
\widetilde{\exists}(V) & = & \sup V\;,\footnotemark
\end{array}
\]
\footnotetext
{We write $\mathcal{P}(S)$, where $S$ is a set, for the powerset of $S$.
}%
where 
$a$, $a_1$, and $a_2$ range over all truth values from $\TValue$ and 
$V$ ranges over all non-empty subsets of $\TValue$.
\end{itemize}
$\TValue$ is the set of \emph{truth values} of $\foBDif(\vSigma)$,
$\DValue$ is the set of \emph{designated truth values} of 
$\foBDif(\vSigma)$, and $\widetilde{\False}$, $\widetilde{\Not}$, 
$\widetilde{\CAnd}$, $\widetilde{\COr}$, $\widetilde{\IImpl}$, 
$\widetilde{\forall}$, and $\widetilde{\exists}$ are the 
\emph{truth functions} that are the interpretations of the logical 
symbols $\False$, $\Not$, $\CAnd$, $\COr$, $\IImpl$, $\forall$, and 
$\exists$, respectively.
The idea behind the designated truth values is that a formula is valid 
if its truth value with respect to all structures and assignments in 
those structures (both defined below) is a designated truth value. 

The set of \emph{non-designated truth values} of $\foBDif(\vSigma)$, 
written $\NDValue$, is the set~$\TValue \diff \DValue$.

\subsubsection*{Structures}

The interpretation of the non-logical symbols of $\foBDif(\vSigma)$ and
the equality symbol is given by means of structures.

A structure $\mathbf{A}$ of $\foBDif(\vSigma)$ is a pair 
$\langle \mathcal{U}\sp\mathbf{A}, \mathcal{I}\sp\mathbf{A} \rangle$, 
where:
\begin{itemize}
\item
$\mathcal{U}\sp\mathbf{A}$ is a set, 
called the \emph{domain of $\mathbf{A}$}, 
such that $\mathcal{U}\sp\mathbf{A} \neq \emptyset$ and
$\mathcal{U}\sp\mathbf{A} \inter \TValue = \emptyset$;
\item
$\mathcal{I}\sp\mathbf{A}$ consists of: 
\pagebreak[2]
\begin{itemize}
\item
an element $c\sp\mathbf{A} \in \mathcal{U}\sp\mathbf{A}$ 
for every $c \in \Func{0}(\vSigma)$;
\item
a function 
$f\sp\mathbf{A}: 
 \overbrace{\mathcal{U}\sp\mathbf{A} \times \cdots \times
             \mathcal{U}\sp\mathbf{A}}^{n+1\; \mathrm{times}} \to
 \mathcal{U}\sp\mathbf{A}$
for every $f \in \Func{n+1}(\vSigma)$ and $n \in \Nat$; 
\item
an element $p\sp\mathbf{A} \in \TValue$ 
for every $p \in \Pred{0}(\vSigma)$;
\item
a function 
$P\sp\mathbf{A}: 
 \overbrace{\mathcal{U}\sp\mathbf{A} \times \cdots \times
             \mathcal{U}\sp\mathbf{A}}^{n+1\; \mathrm{times}} \to
 \TValue$
for every $P \in \Pred{n+1}(\vSigma)$ and $n \in \Nat$;
\item
a function 
$\Meq\sp\mathbf{A}: 
 \mathcal{U}\sp\mathbf{A} \times \mathcal{U}\sp\mathbf{A} \to
 \TValue$
where, for all $d_1,d_2 \in \mathcal{U}\sp\mathbf{A}$,
${\Meq\sp\mathbf{A}}(d_1,d_2) \in \DValue$ iff $d_1 = d_2$.
\end{itemize}
\end{itemize}
Instead of $w\sp\mathbf{A}$ we write $w$ when it is clear from the 
context that the interpretation of symbol $w$ in structure $\mathbf{A}$ 
is meant.

\subsubsection*{Assignments}

The interpretation of the variables of $\foBDif(\vSigma)$ is given by 
means of assignments.

Let $\mathbf{A}$ be a structure of $\foBDif(\vSigma)$.
Then an \emph{assignment in $\mathbf{A}$} is a function
$\alpha: \SVar \to \mathcal{U}\sp\mathbf{A}$.
For every assignment $\alpha$ in $\mathbf{A}$, variable 
$x \in \SVar$, and element $d \in \mathcal{U}\sp\mathbf{A}$, we write 
$\alpha(x \to d)$ for the assignment $\alpha'$ in $\mathbf{A}$ such that 
$\alpha'(x) = d$ and $\alpha'(y) = \alpha(y)$ if $y \not\equiv x$.

\subsubsection*{Valuations and models}

Let $\mathbf{A}$ be a structure of $\foBDif(\vSigma)$, and 
let $\alpha$ be an assignment in $\mathbf{A}$.
Then the \emph{valuation of $\STerm{\vSigma}$ in structure $\mathbf{A}$ 
under assignment $\alpha$} is the func\-tion 
$\mathit{tval}^\mathbf{A}_\alpha :
 \STerm{\vSigma} \to \mathcal{U}\sp\mathbf{A}$ that maps each term $t$ to 
the element of $\mathcal{U}\sp\mathbf{A}$ that is the value of $t$ in 
$\mathbf{A}$ under assignment $\alpha$.
Similarly, the \emph{valuation of $\SForm{\vSigma}$ in structure 
$\mathbf{A}$ under assignment $\alpha$} is the function 
$\mathit{fval}^\mathbf{A}_\alpha :
 \SForm{\vSigma} \to \TValue$ that maps each formula $A$ to the element of 
$\TValue$ that is the truth value of $A$ in $\mathbf{A}$ under assignment 
$\alpha$.
We write $\Term{t}{\mathbf{A}}{\alpha}$ and 
$\Term{A}{\mathbf{A}}{\alpha}$ for 
$\mathit{tval}^\mathbf{A}_\alpha(t)$ and 
$\mathit{fval}^\mathbf{A}_\alpha(A)$, respectively.

These valuation functions are inductively defined in 
Table~\ref{table-interpretation}.
\begin{table}[t!]
\caption{Valuations of terms and formulas of $\foBDif(\vSigma)$}
\label{table-interpretation}
\centering
$
\renewcommand{\arraystretch}{1.25}
\begin{array}[t]{rcl}
\hline
\\[-3ex]
\Term{x}{\mathbf{A}}{\alpha} & = &
 \begin{array}[t]{l}
 \alpha(x) \;,
 \end{array}
\\[.5ex]
\Term{c}{\mathbf{A}}{\alpha} & = &
 \begin{array}[t]{l}
 c\sp\mathbf{A} \;,
 \end{array}
\\[.5ex]
\Term{f(t\sb1,\ldots,t\sb{n+1})}{\mathbf{A}}{\alpha} & = &
 \begin{array}[t]{l}
 f\sp\mathbf{A}(\Term{t\sb1}{\mathbf{A}}{\alpha},\ldots,
                \Term{t\sb{n+1}}{\mathbf{A}}{\alpha})
 \end{array}
\\[1.5ex]
\Term{p}{\mathbf{A}}{\alpha} & = &
 \begin{array}[t]{l}
 p\sp\mathbf{A} \;,
 \end{array}
\\[.5ex]
\Term{P(t\sb1,\ldots,t\sb{n+1})}{\mathbf{A}}{\alpha} & = &
 \begin{array}[t]{l}
 P\sp\mathbf{A}(\Term{t\sb1}{\mathbf{A}}{\alpha},\ldots,
                \Term{t\sb{n+1}}{\mathbf{A}}{\alpha}) \;,
 \end{array}
\\[.5ex]
\Term{t\sb1 = t\sb2}{\mathbf{A}}{\alpha} & = &
 \begin{array}[t]{l}
 \Meq\sp\mathbf{A}(\Term{t\sb1}{\mathbf{A}}{\alpha},
                     \Term{t\sb2}{\mathbf{A}}{\alpha}) \;,
\end{array}
\\[1.5ex]
\Term{\False}{\mathbf{A}}{\alpha} & = & \widetilde{\False} \;,
\\[.5ex]
\Term{\Not A}{\mathbf{A}}{\alpha} & = & 
\widetilde{\Not}(\Term{A}{\mathbf{A}}{\alpha}) \;,
\\[.5ex]
\Term{A\sb1 \CAnd A\sb2}{\mathbf{A}}{\alpha} & = &
\widetilde{\CAnd}(\Term{A\sb1}{\mathbf{A}}{\alpha},
                  \Term{A\sb2}{\mathbf{A}}{\alpha}) \;,
\\[.5ex]
\Term{A\sb1 \COr A\sb2}{\mathbf{A}}{\alpha} & = &
\widetilde{\COr}(\Term{A\sb1}{\mathbf{A}}{\alpha},
                 \Term{A\sb2}{\mathbf{A}}{\alpha}) \;,
\\[.5ex]
\Term{A\sb1 \IImpl A\sb2}{\mathbf{A}}{\alpha} & = &
\widetilde{\IImpl}(\Term{A\sb1}{\mathbf{A}}{\alpha},
                   \Term{A\sb2}{\mathbf{A}}{\alpha}) \;,
\\[.5ex]
\Term{\CForall{x}{A}}{\mathbf{A}}{\alpha} & = &
\widetilde{\forall}(\set{\Term{A}{\mathbf{A}}{\alpha(x \to d)} \where
                         d \in \mathcal{U}\sp\mathbf{A}}) \;,
\\[.5ex]
\Term{\CExists{x}{A}}{\mathbf{A}}{\alpha} & = &
\widetilde{\exists}(\set{\Term{A}{\mathbf{A}}{\alpha(x \to d)} \where
                         d \in \mathcal{U}\sp\mathbf{A}}) \;,
\\[.5ex]
\hline
\end{array}
$
\end{table}
In this table, 
$x$ is a meta-variable ranging over all variables from 
$\SVar$, 
$c$ is a meta-variable ranging over all function symbols from 
$\Func{0}(\vSigma)$,
$f$ is a meta-variable ranging over all function symbols from
$\Func{n+1}(\vSigma)$, 
$p$ is a meta-variable ranging over all predicate symbols from 
$\Pred{0}(\vSigma)$,
$P$ is a meta-variable ranging over all  predicate symbols from
$\Pred{n+1}(\vSigma)$, 
$t_1$, \ldots, $t_{n+1}$ are meta-variables ranging over all terms 
from $\STerm{\vSigma}$, and 
$A$, $A_1$, and $A_2$ are meta-variables ranging over all formulas 
from~$\SForm{\vSigma}$.

The following theorem is a decidability result concerning valuations of
formulas in structures with a finite domain. 
\begin{proposition}
\label{proposition-decidable}
Let $\mathbf{A}$ be a structure of $\foBDif(\vSigma)$ such that
$\mathcal{U}\sp\mathbf{A}$ is finite, and
let $\alpha$ be an  assignment in $\mathbf{A}$.
Then, it is decidable whether, for a formula $A \in \SForm{\vSigma}$,\,  
$\Term{A}{\mathbf{A}}{\alpha} \in \DValue$.
\end{proposition} \pagebreak[2]
\begin{proof}
This is easy to prove by induction on the structure of $A$.
\qed
\end{proof}

Below, the notion of a model of a set of formulas of $\foBDif(\vSigma)$ 
is defined in terms of valuations.

Let $\vGamma$ be a set of formulas from $\SForm{\vSigma}$, and
let $\mathbf{A}$ be a structure of $\foBDif(\vSigma)$.
Then \emph{$\mathbf{A}$ is a model of $\vGamma$} iff
for all assignments $\alpha$ in $\mathbf{A}$, for all $A \in \vGamma$,
$\Term{A}{\mathbf{A}}{\alpha} \in \DValue$.

\subsubsection*{Logical consequence}

Given the valuations of terms and formulas of $\foBDif(\vSigma)$, it is 
easy to make precise when in $\foBDif(\vSigma)$ a set of formulas is a 
logical consequence of another set of formulas.

Let $\vGamma$ and $\vDelta$ be sets of formulas from $\SForm{\vSigma}$.
Then \emph{$\vDelta$ is a logical consequence of $\vGamma$}, written 
$\vGamma \LCon \vDelta$, iff
for all structures $\mathbf{A}$ of $\foBDif(\vSigma)$,
for all assignments $\alpha$ in $\mathbf{A}$,
if $\Term{A}{\mathbf{A}}{\alpha} \in \DValue$ for all $A \in \vGamma$, 
then $\Term{A'}{\mathbf{A}}{\alpha} \in \DValue$ for some 
$A' \in \vDelta$.
We write $\vGamma \notLCon \vDelta$ to indicate that it is not the case 
that $\vGamma \LCon \vDelta$.

The two properties concerning the logical consequence relation of
$\foBDif(\vSigma)$ mentioned below are proved in~\cite{Mid23a}. 

The logical consequence relation $\LCon$ of $\foBDif(\vSigma)$ is such 
that
\[
\vGamma \LCon \vDelta, A_1 \IImpl A_2\;\; \mathrm{iff}\;\; 
A_1, \vGamma \LCon \vDelta, A_2
\]
for all $\vGamma, \vDelta \subseteq \SForm{\vSigma}$ and 
$A_1, A_2 \in \SForm{\vSigma}$ and moreover
there exists a logic $\Logic$ with the same language as 
$\foBDif(\vSigma)$ and a logical consequence relation $\LCon'$ such that:
\begin{itemize}
\item
${\LCon} \subseteq {\LCon'}$;
\item
the matrix $\langle \TValue', \DValue', \TFunct' \rangle$ of $\Logic$ is 
such that $\TValue' = \set{\VTrue,\VFalse}$, $\DValue' = \set{\VTrue}$, 
and the interpretation of $\Not$ in $\TFunct'$ is as follows:
\[
\renewcommand{\arraystretch}{1.25}
\begin{array}[t]{c}
\widetilde{\Not}'(a) =
 \left \{
 \begin{array}{l@{\;\;}l}
 \VTrue  & \mathrm{if}\; a = \VFalse \\
 \VFalse & \mathrm{if}\; a = \VTrue
 \end{array}
 \right.
\end{array}
\]
for all $a \in \TValue'$.
\end{itemize}
This means that $\foBDif(\vSigma)$ is $\Not$-\emph{coherent with 
classical logic} in the sense of~\cite{AA17a}.

There exist $A, A' \in \SForm{\vSigma}$ such that 
$A, \Not A \notLCon A'$.
Because $\foBDif(\vSigma)$ is also $\Not$-coherent with classical logic, 
this means that $\foBDif(\vSigma)$ is \emph{paraconsistent} in the sense 
of~\cite{AA17a}.

There exist a $\vGamma \subseteq \SForm{\vSigma}$ and
$A, A' \in \SForm{\vSigma}$ such that $\vGamma, A \LCon A'$ and 
$\vGamma, \Not A \LCon A'$, but $\vGamma \notLCon A'$.
Because $\foBDif(\vSigma)$ is also $\Not$-coherent with classical logic, 
this means that $\foBDif(\vSigma)$ is \emph{paracomplete} in the sense 
of~\cite{AA17a}.

\subsubsection*{Abbreviations}

In what follows, the following abbreviations will be used: 
\[
\renewcommand{\arraystretch}{1.25}
\begin{array}[t]{r@{\;\;}c@{\;\;}l@{\;\;}l}
t_1 \neq t_2   & \mathrm{stands\; for} & \Not (t_1 = t_2), \\
\True          & \mathrm{stands\; for} & \Not \False, \\
A_1 \SImpl A_2 & \mathrm{stands\; for} &
              (A_1 \IImpl A_2) \CAnd (\Not A_2 \IImpl \Not A_1), \\
\Des A         & \mathrm{stands\; for} & \Not (A \IImpl \False) &
\mathrm{(designatedness)}, \\
\Cons A        & \mathrm{stands\; for} & \Not (\Des (A \CAnd \Not A)) &
\mathrm{(consistency)}, \\
\Det A         & \mathrm{stands\; for} & \Des (A \COr \Not A) &
\mathrm{(determinacy)}.
\end{array}
\]
It follows from the definitions concerned that:
\[
\begin{array}[t]{l@{\;\;}c@{\;\;}l}
\Term{\Des A}{\mathbf{A}}{\alpha} & = &
 \left \{
 \begin{array}{l@{\;\;}l}
 \VTrue  & \mathrm{if}\; 
           \Term{A}{\mathbf{A}}{\alpha} \in \set{\VTrue,\VBoth} \\
 \VFalse & \mathrm{otherwise},
 \end{array}
 \right.
\vspace*{.5ex} \\
\Term{\Cons A}{\mathbf{A}}{\alpha} & = &
 \left \{
 \begin{array}{l@{\;\;}l}
 \VTrue  & \mathrm{if}\; 
      \Term{A}{\mathbf{A}}{\alpha} \in \set{\VTrue,\VFalse,\VNeither} \\
 \VFalse & \mathrm{otherwise},
 \end{array}
 \right.
\vspace*{.5ex} \\
\Term{\Det A}{\mathbf{A}}{\alpha} & = &
 \left \{
 \begin{array}{l@{\;\;}l}
 \VTrue  & \mathrm{if}\; 
         \Term{A}{\mathbf{A}}{\alpha} \in \set{\VTrue,\VFalse,\VBoth} \\
 \VFalse & \mathrm{otherwise}.
 \end{array}
 \right.
\end{array}
\]
This means that the abbreviations $\Des A$, $\Cons A$, and $\Det A$ 
correspond to formulas of studied expansions of Belnap-Dunn logic whose 
connectives include $\Des$, $\Cons$ or $\Det$.
The connective $\Des$ is for example found in the expansion of 
Belnap-Dunn logic known as BD$\mathrm{\vDelta}$~\cite{SO14a}.
The connectives $\Cons$ and $\Det$ have for example been studied in the 
setting of Belnap-Dunn logic in~\cite{CC20a}.
The connective $\Cons$ is also found in the expansion of the first-order 
version of Priest's logic of paradox~\cite{Pri79a} known as 
\foLPc~\cite{Pic18a}.
The connective $\Det$ is the counterpart of $\Cons$ in the setting of 
Kleene's strong three-valued logic~\cite[Section~64]{Kle52a}. 

Moreover, notice that:
\[
\begin{array}[t]{l@{\;\;}c@{\;\;}l}
\Term{\Cons A \CAnd \Det A}{\mathbf{A}}{\alpha} & = &
 \left \{
 \begin{array}{l@{\;\;}l}
 \VTrue  & \mathrm{if}\; 
         \Term{A}{\mathbf{A}}{\alpha} \in \set{\VTrue,\VFalse} \\
 \VFalse & \mathrm{otherwise}.
 \end{array}
 \right.
\end{array}
\]

\section{A Proof System for $\foBDif(\vSigma)$}
\label{PROOF-RULES}

In this section, a sequent calculus proof system for $\foBDif(\vSigma)$ 
is presented.
This means that the inference rules have sequents as premises and 
conclusions.
First, the notion of a sequent is introduced. 
Then, the inference rules of the proof system of $\foBDif(\vSigma)$ are 
presented.
After that, the notion of a derivation of a sequent from a set of 
sequents and the notion of a proof of a sequent are introduced.
Extensions of the proof system of $\foBDif(\vSigma)$ which can serve as 
proof system for closely related logic are also described.

\subsubsection*{Sequents}

In the sequent calculus proof system for $\foBDif(\vSigma)$, a 
\emph{sequent} is an expression of the form $\vGamma \scEnt \vDelta$, 
where $\vGamma$ and $\vDelta$ are finite sets of formulas from 
$\SForm{\vSigma}$.
We write $\vGamma,\vGamma'$ for $\vGamma \union \vGamma'$ and 
$A$, where $A$ is a formula from $\SForm{\vSigma}$, for $\set{A}$ on both 
sides of a sequent.
Moreover, we write ${} \scEnt \vDelta$ instead of 
$\emptyset \scEnt \vDelta$. 

A sequent $\vGamma \scEnt \vDelta$ states that the logical consequence 
relation that is defined in Section~\ref{INTERPRETATION} holds between 
$\vGamma$ and $\vDelta$.
If a sequent $\vGamma \scEnt \vDelta$ can be proved by means of the rules 
of inference given below, then that logical consequence relation holds 
between $\vGamma$ and~$\vDelta$.

\subsubsection*{Rules of inference}

The sequent calculus proof system for $\foBDif(\vSigma)$ consists of the 
inference rules given in Table~\ref{table-proof-system}.
\begin{table}[!p]
\caption{A sequent calculus proof system for $\foBDif$}
\label{table-proof-system}
\renewcommand{\arraystretch}{1.1} 
\centering 
\begin{small}
\begin{tabular}[t]{@{}c@{}}
\hline
\\[-2.5ex]
\begin{tabular}{@{}l@{}}
\InfRuleC{Id}
 {{}}
 {A, \vGamma \scEnt \vDelta, A}
 {$\ast$}
\\[2.5ex] 
\InfRule{$\False$-L}
 {{}}
 {\False, \vGamma \scEnt \vDelta}
\\[2.5ex] 
\InfRule{$\CAnd$-L}
 {A\sb1, A\sb2, \vGamma \scEnt \vDelta}
 {A\sb1 \CAnd A\sb2, \vGamma \scEnt \vDelta}
\\[2.5ex]
\InfRule{$\COr$-L}
 {A\sb1, \vGamma \scEnt \vDelta \quad
  A\sb2, \vGamma \scEnt \vDelta}
 {A\sb1 \COr A\sb2, \vGamma \scEnt \vDelta}
\\[2.5ex]
\InfRule{$\IImpl$-L}
 {\vGamma \scEnt \vDelta, A\sb1 \quad
  A\sb2, \vGamma \scEnt \vDelta}
 {A\sb1 \IImpl A\sb2, \vGamma \scEnt \vDelta}
\\[2.5ex]
\InfRule{$\forall$-L}
 {\subst{x \assign t}A, \vGamma \scEnt \vDelta}
 {\CForall{x}{A}, \vGamma \scEnt \vDelta}
\\[2.5ex]
\InfRuleC{$\exists$-L}
 {\subst{x \assign y}A, \vGamma \scEnt \vDelta}
 {\CExists{x}{A}, \vGamma \scEnt \vDelta}
 {\dag}
\\[2.5ex]
\phantom{
\InfRule{$\Not \False$-R}
 {{}}
 {\vGamma \scEnt \vDelta, \Not \False}
}
\\[2.5ex] 
\InfRule{$\Not \Not$-L}
 {A, \vGamma \scEnt \vDelta}
 {\Not \Not A, \vGamma \scEnt \vDelta}
\\[2.5ex]
\InfRule{$\Not \CAnd$-L}
 {\Not A\sb1, \vGamma \scEnt \vDelta \quad
  \Not A\sb2, \vGamma \scEnt \vDelta}
 {\Not (A\sb1 \CAnd A\sb2), \vGamma \scEnt \vDelta}
\\[2.5ex]
\InfRule{$\Not \COr$-L}
 {\Not A\sb1, \Not A\sb2, \vGamma \scEnt \vDelta}
 {\Not (A\sb1 \COr A\sb2), \vGamma \scEnt \vDelta}
\\[2.5ex]
\InfRule{$\Not \IImpl$-L}
 {A\sb1, \Not A\sb2, \vGamma \scEnt \vDelta}
 {\Not (A\sb1 \IImpl A\sb2), \vGamma \scEnt \vDelta}
\\[2.5ex]
\InfRuleC{$\Not \forall$-L}
 {\Not \subst{x \assign y}A, \vGamma \scEnt \vDelta}
 {\Not \CForall{x}{A}, \vGamma \scEnt \vDelta}
 {\dag}
\\[2.5ex]
\InfRule{$\Not \exists$-L}
 {\Not \subst{x \assign t}A, \vGamma \scEnt \vDelta}
 {\Not \CExists{x}{A}, \vGamma \scEnt \vDelta}
\\[2.5ex]
\InfRule{$=$-Refl}
 {t = t, \vGamma \scEnt \vDelta}
 {\vGamma \scEnt \vDelta}
\\[2.5ex]
\end{tabular}
\qquad 
\begin{tabular}{@{}l@{}} 
\InfRule{Cut}
 {\vGamma \scEnt \vDelta, A \quad
  A, \vGamma' \scEnt \vDelta'}
 {\vGamma', \vGamma \scEnt \vDelta, \vDelta'}
\\[2.5ex] 
\phantom{
\InfRule{$\False$-L}
 {{}}
 {\False, \vGamma \scEnt \vDelta}
}
\\[2.5ex]
\InfRule{$\CAnd$-R}
 {\vGamma \scEnt \vDelta, A\sb1 \quad
  \vGamma \scEnt \vDelta, A\sb2}
 {\vGamma \scEnt \vDelta, A\sb1 \CAnd A\sb2}
\\[2.5ex] 
\InfRule{$\COr$-R}
 {\vGamma \scEnt \vDelta, A\sb1, A\sb2}
 {\vGamma \scEnt \vDelta, A\sb1 \COr A\sb2}
\\[2.5ex]
\InfRule{$\IImpl$-R}
 {A\sb1, \vGamma \scEnt \vDelta, A\sb2}
 {\vGamma \scEnt \vDelta, A\sb1 \IImpl A\sb2}
\\[2.5ex]
\InfRuleC{$\forall$-R}
 {\vGamma \scEnt \vDelta, \subst{x \assign y}A}
 {\vGamma \scEnt \vDelta, \CForall{x}{A}}
 {\dag}
\\[2.5ex]
\InfRule{$\exists$-R}
 {\vGamma \scEnt \vDelta, \subst{x \assign t}A}
 {\vGamma \scEnt \vDelta, \CExists{x}{A}}
\\[2.5ex]
\InfRule{$\Not \False$-R}
 {{}}
 {\vGamma \scEnt \vDelta, \Not \False}
\\[2.5ex] 
\InfRule{$\Not \Not$-R}
 {\vGamma \scEnt \vDelta, A}
 {\vGamma \scEnt \vDelta, \Not \Not A}
\\[2.5ex] 
\InfRule{$\Not \CAnd$-R}
 {\vGamma \scEnt \vDelta, \Not A\sb1, \Not A\sb2}
 {\vGamma \scEnt \vDelta, \Not (A\sb1 \CAnd A\sb2)}
\\[2.5ex] 
\InfRule{$\Not \COr$-R}
 {\vGamma \scEnt \vDelta, \Not A\sb1 \quad
  \vGamma \scEnt \vDelta, \Not A\sb2}
 {\vGamma \scEnt \vDelta, \Not (A\sb1 \COr A\sb2)}
\\[2.5ex] 
\InfRule{$\Not \IImpl$-R}
 {\vGamma \scEnt \vDelta, A\sb1 \quad
  \vGamma \scEnt \vDelta, \Not A\sb2}
 {\vGamma \scEnt \vDelta, \Not (A\sb1 \IImpl A\sb2)}
\\[2.5ex]
\InfRule{$\Not \forall$-R}
 {\vGamma \scEnt \vDelta, \Not \subst{x \assign t}A}
 {\vGamma \scEnt \vDelta, \Not \CForall{x}{A}}
\\[2.5ex]
\InfRuleC{$\Not \exists$-R}
 {\vGamma \scEnt \vDelta, \Not \subst{x \assign y}A}
 {\vGamma \scEnt \vDelta, \Not \CExists{x}{A}}
 {\dag}
\\[2.5ex]
\InfRuleC{$=$-Repl}
 {\subst{x \assign t\sb1}A, \vGamma \scEnt \vDelta}
 {t\sb1 = t\sb2, \subst{x \assign t\sb2}A, \vGamma \scEnt \vDelta}
 {$\ast$}
\\[2.5ex]
\end{tabular}
\\[2.5ex]
\begin{tabular}{@{}l@{}}
$\ast$ restriction: $A$ is a literal.
\\ 
$\dag$ restriction:
$y$ is not free in $\vGamma$, $y$ is not free in $\vDelta$,
$y$ is not free in $A$ unless $x \equiv y$.
\vspace*{1ex} \par
\end{tabular}
\\
\hline
\end{tabular}
\end{small}
\end{table}
In this table, 
$x$ and $y$ are meta-variables ranging over all variables 
from $\SVar$,
$t$, $t_1$, and $t_2$ are meta-variables ranging over all terms 
from $\STerm{\vSigma}$, 
$A$, $A_1$, and $A_2$ are meta-variables ranging over all formulas 
from $\SForm{\vSigma}$, and
$\vGamma$ and $\vDelta$ are meta-variables ranging over all finite 
sets of formulas from $\SForm{\vSigma}$. 
A formula $A$ from $\SForm{\vSigma}$ is a \emph{literal} iff
$A \in \SLit{\vSigma}$.

\subsubsection*{Derivations and proofs}

In the sequent calculus proof system for $\foBDif(\vSigma)$, 
a \emph{derivation of a sequent $\vGamma \scEnt \vDelta$ from a finite set 
of sequents $\mathcal{H}$} is a finite sequence $\seq{s_1,\ldots,s_n}$ 
of sequents such that $s_n$ equals $\vGamma \scEnt \vDelta$ and, for each 
$i \in \set{1,\ldots,n}$, one of the following conditions holds:
\begin{itemize}
\item
$s_i \in \mathcal{H}$;
\item
$s_i$ is the conclusion of an instance of some inference rule from the 
proof system of $\foBDif(\vSigma)$ whose premises are among 
$s_1,\ldots,s_{i-1}$.
\end{itemize}
A \emph{proof of a sequent $\vGamma \scEnt \vDelta$} is a derivation of 
$\vGamma \scEnt \vDelta$ from the empty set of sequents.
A sequent $\vGamma \scEnt \vDelta$ is said to be \emph{provable} if there 
exists a proof of $\vGamma \scEnt \vDelta$.

Let $\vGamma$ and $\vDelta$ be sets of formulas from $\SForm{\vSigma}$.
Then $\vDelta$ is \emph{derivable} from $\vGamma$, written 
$\vGamma \LDer \vDelta$, iff there exist finite sets 
$\vGamma' \subseteq \vGamma$ and $\vDelta' \subseteq \vDelta$ such that the 
sequent $\vGamma' \scEnt \vDelta'$ is provable.

The sequent calculus proof system of $\foBDif(\vSigma)$ is sound and 
complete with respect to the logical consequence relation $\LCon$ 
defined in Section~\ref{INTERPRETATION}.
\begin{theorem}
\label{theorem-sound-complete}
Let $\vGamma$ and $\vDelta$ be sets of formulas from $\SForm{\vSigma}$. 
Then $\vGamma \LDer \vDelta$ iff $\vGamma \LCon \vDelta$.
\end{theorem}
\begin{proof}
See Appendix~A of~\cite{Mid23a}.
\qed
\end{proof}

\subsubsection*{Extensions of the proof system}

The languages of $\FOCL$ and $\foBDif$ are the same.
A sound and complete sequent calculus proof system of $\FOCL$ 
can be obtained by adding the following two inference rules to the 
sequent calculus proof system of $\foBDif$: 
\\[1.5ex]
\mbox{} \hfill
\begin{tabular}{@{}c@{}} 
\InfRule{$\Not$-L}
 {\vGamma \scEnt \vDelta, A}
 {\Not A, \vGamma \scEnt \vDelta}
\hspace*{4em}
\InfRule{$\Not$-R}
 {A, \vGamma \scEnt \vDelta}
 {\vGamma \scEnt \vDelta, \Not A}
\end{tabular}
\hfill
\vspace*{1.5ex}

If we add only the inference rule $\Not$-R to the sequent calculus proof 
system of $\foBDif$, then we obtain a sound and complete proof 
system of the paraconsistent (but not paracomplete) logic \foLPif\ 
presented in~\cite{Mid22b}.
If we add only the inference rule $\Not$-L to the sequent calculus proof 
system of $\foBDif$, then we obtain a sound and complete proof 
system of the obvious first-order version of the paracomplete (but not 
paraconsistent) propositional logic $\Klif$ presented in~\cite{Mid17a}.

\section{Covering Terms with an Indeterminate Value}
\label{INDETERMINATE}

This section goes into a minor variation of $\foBDif(\vSigma)$, called
$\pfoBDif(\vSigma)$, that can deal with terms with an indeterminate value.
Semantically, this is handled by restricting the domain of structures to 
sets that contain a special dummy value $\bot$ and assigning this value 
to terms with an indeterminate value.
Thus, $\bot$ is treated as a genuine value.
It differs from other values only by conditions imposed on the equality 
relation of structures (see below).
The conditions concerned fit in with the intuition that a term with an 
indeterminate value has an unknown value or does not have a value.

The language of $\foBDif(\vSigma)$ and the language of $\pfoBDif(\vSigma)$ 
are the same.
The logical consequence relation of $\pfoBDif(\vSigma)$ is defined as for
$\foBDif(\vSigma)$, but in terms of structures of $\pfoBDif(\vSigma)$.
These structures differ slightly from the structures of 
$\foBDif(\vSigma)$.  

\subsubsection*{Structures}

A \emph{structure} $\mathbf{A}$ of $\pfoBDif(\vSigma)$ is a pair 
$\langle \mathcal{U}\sp\mathbf{A}, \mathcal{I}\sp\mathbf{A} \rangle$, 
where:
\begin{itemize}
\item
$\mathcal{U}\sp\mathbf{A}$ is a set, 
called the \emph{domain of $\mathbf{A}$}, 
such that $\bot \in \mathcal{U}\sp\mathbf{A}$,\;
$\mathcal{U}\sp\mathbf{A} \diff \set{\bot} \neq \emptyset$, and
$\mathcal{U}\sp\mathbf{A} \inter \TValue = \emptyset$;
\item
$\mathcal{I}\sp\mathbf{A}$ consists of: 
\begin{itemize}
\item
an element $c\sp\mathbf{A} \in \mathcal{U}\sp\mathbf{A}$ 
for every $c \in \Func{0}(\vSigma)$;
\item
a function 
$f\sp\mathbf{A}: 
 \overbrace{\mathcal{U}\sp\mathbf{A} \times \cdots \times
             \mathcal{U}\sp\mathbf{A}}^{n+1\; \mathrm{times}} \to
 \mathcal{U}\sp\mathbf{A}$
for every $f \in \Func{n+1}(\vSigma)$ and $n \in \Nat$; 
\item
an element $p\sp\mathbf{A} \in \TValue$ 
for every $p \in \Pred{0}(\vSigma)$;
\item
a function 
$P\sp\mathbf{A}: 
 \overbrace{\mathcal{U}\sp\mathbf{A} \times \cdots \times
             \mathcal{U}\sp\mathbf{A}}^{n+1\; \mathrm{times}} \to
 \TValue$
for every $P \in \Pred{n+1}(\vSigma)$ and $n \in \Nat$;
\item
a function 
$\Meq\sp\mathbf{A}: 
 \mathcal{U}\sp\mathbf{A} \times \mathcal{U}\sp\mathbf{A} \to
 \TValue$
where, for all $d_1,d_2 \in \mathcal{U}\sp\mathbf{A}$:
\begin{itemize}
\item[]
${\Meq\sp\mathbf{A}}(d_1,d_2) \in \DValue$ iff 
$d_1,d_2 \in \mathcal{U}\sp\mathbf{A} \diff \set{\bot}$ and $d_1 = d_2$;
\item[]
${\Meq\sp\mathbf{A}}(d_1,d_2) = \VNeither\;$ iff 
$d_1 = \bot$ or $d_2 = \bot$.
\end{itemize}
\end{itemize}
\end{itemize}

By the first condition imposed on ${\Meq\sp\mathbf{A}}$ in the 
definition of a structure $\mathbf{A}$ of $\pfoBDif(\vSigma)$,
$\Term{t = t}{\mathbf{A}}{\alpha} \in \DValue$ iff 
$\Term{t}{\mathbf{A}}{\alpha} \neq \bot$.
In other words, the truth value of the formula $t = t$ is a designated 
truth value iff the value of the term $t$ is determinate. 

By the second condition imposed on ${\Meq\sp\mathbf{A}}$ in the 
definition of a structure $\mathbf{A}$ of $\pfoBDif(\vSigma)$,
$\Term{t\sb1 = t\sb2}{\mathbf{A}}{\alpha} = \VNeither$ iff 
$\Term{t\sb1}{\mathbf{A}}{\alpha} = \bot$ or
$\Term{t\sb2}{\mathbf{A}}{\alpha} = \bot$.
In other words, the truth value of an equation $t\sb1 = t\sb2$ is 
neither true nor false~($\VNeither$) iff the value of $t\sb1$ or $t\sb2$ 
or both is indeterminate.
Similar conditions are not imposed on the interpretations of the 
function and predicate symbols from $\vSigma$. 
However, because of the first condition imposed on 
${\Meq\sp\mathbf{A}}$, the relevant conditions can be expressed in 
$\pfoBDif(\vSigma)$.

\subsubsection*{Logical consequence}

Let $\vGamma$ and $\vDelta$ be sets of formulas from $\SForm{\vSigma}$.
Then \emph{$\vDelta$ is a logical consequence of $\vGamma$}, written 
$\vGamma \mathrel{\LCon_\bot} \vDelta$, iff
for all structures $\mathbf{A}$ of $\pfoBDif(\vSigma)$,
for all assignments $\alpha$ in $\mathbf{A}$,
if $\Term{A}{\mathbf{A}}{\alpha} \in \DValue$ for all $A \in \vGamma$, 
then $\Term{A'}{\mathbf{A}}{\alpha} \in \DValue$ for some 
$A' \in \vDelta$.

By the conditions on ${\Meq\sp\mathbf{A}}$ in the definition of a 
structure $\mathbf{A}$ of $\pfoBDif(\vSigma)$, we have that, for all 
$t_1,t_2 \in \STerm{\vSigma}$: 
\[
\renewcommand{\arraystretch}{1.25}
\begin{array}{@{}c@{}}
t_1 = t_1 \CAnd t_2 = t_2 \mathrel{\LCon_\bot}
t_1 = t_2 \COr \Not (t_1 = t_2)\;, 
\\
t_1 = t_2 \COr \Not (t_1 = t_2) \mathrel{\LCon_\bot}
t_1 = t_1 \CAnd t_2 = t_2\;.
\end{array}
\]

\subsubsection*{Proof system}

A sequent calculus proof system of $\pfoBDif(\vSigma)$ is obtained by 
replacing the inference rule $=$-Refl by the following two inference 
rules in the sequent calculus proof system of $\foBDif(\vSigma)$: 
\[
\begin{tabular}{@{}l@{}} 
\InfRule{$\delta$-$=$-L}
 {t_1 = t_1 \CAnd t_2 = t_2, \vGamma \scEnt \vDelta}
 {t_1 = t_2 \COr t_1 \neq t_2, \vGamma \scEnt \vDelta}
\\[3ex]
\InfRule{$\delta$-$=$-R}
 {\vGamma \scEnt \vDelta, t_1 = t_1 \CAnd t_2 = t_2}
 {\vGamma \scEnt \vDelta, t_1 = t_2 \COr t_1 \neq t_2} 
\end{tabular}
\]
The resulting proof system is sound and complete.
It is easy to see that the resulting proof system is sound.
As explained in~\cite{Mid23a}, the completeness proof requires a minor 
adaptation of the completeness proof for the proof system of 
$\foBDif(\vSigma)$ given in that paper. 

\subsubsection*{Abbreviations}

In what follows, the following abbreviations will be used: 
\[
\renewcommand{\arraystretch}{1.25}
\begin{array}[t]{r@{\;\;}c@{\;\;}l@{\;\;}l}
t \Def       & \mathrm{stands\; for} & \Des (t = t) &
\mathrm{(term\; determinacy)}, \\
t_1 \SEq t_2 & \mathrm{stands\; for} & 
               t_1 = t_2 \COr \Not (t_1 \Def \COr t_2 \Def) &
\mathrm{(strong\; equality)}.
\end{array}
\]
It follows from these definitions that:
\[
\begin{array}[t]{l@{\;\;}c@{\;\;}l}
\Term{t \Def}{\mathbf{A}}{\alpha} & = &
 \left \{
 \begin{array}{l@{\;\;}l}
 \VTrue  & \mathrm{if}\; 
           \Term{t}{\mathbf{A}}{\alpha} \in
           \mathcal{U}\sp\mathbf{A} \diff \set{\bot} \\
 \VFalse & \mathrm{otherwise},
 \end{array}
 \right.
\vspace*{.5ex} \\
\Term{t_1 \SEq t_2}{\mathbf{A}}{\alpha} & = &
 \left \{
 \begin{array}{l@{\;\;}l}
 \VTrue  & \mathrm{if}\; 
           \Term{t\sb1}{\mathbf{A}}{\alpha} = \bot\; \mathrm{and}\;
           \Term{t\sb2}{\mathbf{A}}{\alpha} = \bot \\
 \Meq\sp\mathbf{A}(\Term{t\sb1}{\mathbf{A}}{\alpha},
                   \Term{t\sb2}{\mathbf{A}}{\alpha})
         & \mathrm{otherwise}.
 \end{array}
 \right.
\end{array}
\]

\subsubsection*{Notes}

$\pfoBDif(\vSigma)$ looks like a free logic~\cite{Lam67a,Nol07a} that is 
paraconsistent and paracomplete.
However, $\bot$ is included in the range of variables in 
$\pfoBDif(\vSigma)$, whereas $\bot$ would be excluded from the range of 
variables in a free logic. 
This difference reflects that~$\bot$ is treated as a genuine value in 
$\pfoBDif(\vSigma)$ and $\bot$ is not treated as a genuine value in 
free logics.

$\pfoBDif(\vSigma)$ has been devised to deal with logical theories 
(i.e.\ sets of formulas) of the kind that arise for example when 
possibly inconsistent relational databases with possibly null values are 
viewed as logical theories.

\section{Relational Databases Viewed through \pfoBDif}
\label{DATABASE}

In this section, relational databases are considered from the 
perspective of \pfoBDif, taking into account that a database may be an 
inconsistent database (cf.~\cite{Bry97a,ABC99a}) and/or a database with 
null values (cf.~\cite{Vas79a,Zan84a,IL84a}).
The proof theoretic point of view is taken, i.e.\ a relational database 
is viewed as a logical theory. 
In the definition of the notion of a relational database, use is made of
the notions of a relational language and a relational theory.
The latter two notions are defined first.
The definitions given in this section are to a great extent based on 
those given in~\cite{Rei84a}.
However, types are ignored for the sake of simplicity 
(cf.~\cite{GMN84a,Var86a}). 
The view on null values in relational databases taken here is informally 
described in Section~\ref{INTRO}.

\subsubsection*{Relational languages}

The pair $(\vSigma,\SForm{\vSigma})$, where $\vSigma$ is a signature, is 
called the \emph{language of $\pfoBDif(\vSigma)$}.
If $\vSigma$ satisfies particular conditions, then the language of 
$\pfoBDif(\vSigma)$ is considered a relational language.

Let $\vSigma$ be a signature.
Then the language $R = (\vSigma,\SForm{\vSigma})$ of $\pfoBDif(\vSigma)$ is 
a \emph{relational language} iff it satisfies the following conditions:
\begin{itemize} 
\item
$\Func{0}(\vSigma)$ is finite,
$\Null \in \Func{0}(\vSigma)$, and
$\Func{0}(\vSigma) \diff \set{\Null}$ is non-empty;
\item
$\Union \set{\Func{n+1}(\vSigma) \where n \in \Nat}$ is empty;
\item
$\Pred{0}(\vSigma)$ is empty;
\item
$\Union \set{\Pred{n+1}(\vSigma) \where n \in \Nat}$ is finite.
\end{itemize}

\subsubsection*{Relational theories}

Below, we will introduce the notion of a relational theory.
In the definition of a relational theory, use is made of a number of 
auxiliary notions.
These auxiliary notions are defined first.

Let $R = (\vSigma,\SForm{\vSigma})$ be a relational language.
Then an \emph{atomic fact for $R$} is a formula from $\SForm{\vSigma}$ of 
the form $P(c_1,\ldots,c_{n+1})$, where $P \in \Pred{n+1}(\vSigma)$ and 
$c_1,\ldots,c_{n+1} \in \Func{0}(\vSigma)$. 

Let $R = (\vSigma,\SForm{\vSigma})$ be a relational language.
Then the \emph{nil-indeterminacy axiom for $R$} is the formula
\[
\Not (\Null \Def)
\]
and the \emph{equality semi-normality axiom for $R$} is the formula
\[
\CForall{x,x'}
 {\Cons (x = x') \CAnd ((x \Def \CAnd x' \Def) \SImpl \Det (x = x'))}\;.
\]

Let $R = (\vSigma,\SForm{\vSigma})$ be a relational language and
let $c_1,\ldots,c_n$ be all members of $\Func{0}(\vSigma)$.
Then the \emph{domain closure axiom for $R$} is the formula
\[
\CForall{x}{(x \SEq c_1 \COr \ldots \COr x \SEq c_n})\;
\]
and the \emph{unique name axiom set for $R$} is the set of formulas
\[
\set{\Not (c_i \SEq c_j) \where 1 \leq i < j \leq n}\;.
\]

Let $R = (\vSigma,\SForm{\vSigma})$ be a relational language and 
let $P \in \Pred{n+1}(\vSigma)$ ($n \in \Nat$). 
Then the \emph{$P$-determinacy axiom for $R$} is the formula
\[
\CForall{x_1,\ldots,x_{n+1}}{\Det P(x_1,\ldots,x_{n+1})}\;.
\]

Let $R = (\vSigma,\SForm{\vSigma})$ be a relational language,
let $\vLambda \subseteq \SForm{\vSigma}$ be a finite set of atomic facts
for $R$, and let $P \in \Pred{n+1}(\vSigma)$ ($n \in \Nat$). 
Suppose that there exist formulas in $\vLambda$ in which $P$ occurs and
let $P(c^1_1,\ldots,c^1_{n+1})$, \ldots, $P(c^m_1,\ldots,c^m_{n+1})$ be 
all formulas from $\vLambda$ in which $P$ occurs.
Then the \emph{$P$-completion axiom for $\vLambda$} is the formula
\[
\begin{array}[t]{@{}l@{}}
\CForall{x_1,\ldots,x_{n+1}}
 {P(x_1,\ldots,x_{n+1}) \SImpl {} \\ \hspace*{.75em} 
  x_1 \SEq c^1_1 \CAnd \ldots \CAnd x_{n+1} \SEq c^1_{n+1}
   \,\COr\, \ldots \,\COr\,
  x_1 \SEq c^m_1 \CAnd \ldots \CAnd x_{n+1} \SEq c^m_{n+1}}\;.
\end{array}
\]
Suppose that there does not exist a formula in $\vLambda$ in which $P$ 
occurs.
Then the \emph{$P$-completion axiom for $\vLambda$} is the formula
\[
\begin{array}[t]{@{}l@{}}
\CForall{x_1,\ldots,x_{n+1}}{P(x_1,\ldots,x_{n+1}) \SImpl \False}\;.
\end{array}
\]

The domain closure, unique name, and $P$-completion axioms are adopted 
from~\cite{Rei84a}.
The nil-indeterminacy, equality semi-normality, and $P$-determinacy 
axioms are new.
The nil-indeterminacy axiom states that the value of $\Null$ is  
indeterminate.
The equality semi-normality axiom states that equations are interpreted 
classically except that their truth value is neither true nor false if 
terms with an indeterminate value are involved.
The $P$-determinacy axiom states that $P$ never yields the truth value
neither true nor false.

Let $R = (\vSigma,\SForm{\vSigma})$ be a relational language.
Then the \emph{relational structure axioms for $R$}, written
$\mathit{RSA}(R)$, is the set of all formulas 
$A \in \SForm{\vSigma}$ for which one of the following holds:
\begin{itemize} 
\item
$A$ is the nil-indeterminacy axiom for $R$;
\item
$A$ is the equality semi-normality axiom for $R$;
\item
$A$ is the domain closure axiom for $R$;
\item
$A$ is an element of the unique name axiom set for $R$;
\item
$A$ is the $P$-determinacy axiom for $R$ 
for some $P \in \Union \set{\Pred{n+1}(\vSigma) \where n \in \Nat}$.
\end{itemize} 

Let $R = (\vSigma,\SForm{\vSigma})$ be a relational language, and
let $\vLambda \subseteq \SForm{\vSigma}$ be a finite set of atomic facts
for $R$.
Then the \emph{relational theory for $R$ with basis $\vLambda$}, written
$\mathit{RT}(R,\vLambda)$, is the set of all formulas 
$A \in \SForm{\vSigma}$ for which one of the following holds:
\begin{itemize} 
\item
$A \in \mathit{RSA}(R)$;
\item
$A \in \vLambda$;
\item
$A$ is the $P$-completion axiom for $\vLambda$
for some $P \in \Union \set{\Pred{n+1}(\vSigma) \where n \in \Nat}$.
\end{itemize}
A set $\vTheta \subseteq \SForm{\vSigma}$ is called a 
\emph{relational theory for $R$} if $\vTheta = \mathit{RT}(R,\vLambda)$ 
for some finite set $\vLambda \subseteq \SForm{\vSigma}$ of atomic facts 
for $R$.
The elements of this unique $\vLambda$ are called the 
\emph{atomic facts of $\vTheta$}.

The following theorem is a decidability result concerning provability 
of sequents $\vGamma \scEnt A$ where $\vGamma$ includes the relational 
structure axioms for some relational language.
\begin{theorem}
\label{theorem-decidable-add}
Let $R = (\vSigma,\SForm{\vSigma})$ be a relational language, and
let $\vGamma$ be a finite subset of $\SForm{\vSigma}$ such that 
$\mathit{RSA}(R) \subseteq \vGamma$.
Then it is decidable whether, for a formula $A \in \SForm{\vSigma}$,\, 
$\vGamma \scEnt A$ is provable. 
\end{theorem}
\begin{proof}
Since it is known from Theorem~\ref{theorem-sound-complete} that 
$\vGamma \scEnt A$ is provable iff $\vGamma \LCon A$, it is shown instead 
that it is decidable whether, for a formula $A \in \SForm{\vSigma}$,\, 
$\vGamma \LCon A$.

Because $\mathit{RSA}(R) \subseteq \vGamma$, it is sufficient to consider 
only structures that are models of $\mathit{RSA}(R)$.
The domains of these structures have the same finite cardinality. 
Because in addition there are finitely many predicate symbols in 
$\vSigma$, there exist moreover only finitely many of these structures.

Clearly, it is sufficient to consider only the restrictions of 
assignments to the set of all variables occurring in $\vGamma \union \set{A}$.
Because the set of all variable occurring in $\vGamma \union \set{A}$ 
is finite and the domain of the structures to be considered is finite, 
there exist only finitely many such restrictions and those restrictions 
are finite.

It follows easily from the above-mentioned finiteness properties and Proposition~\ref{proposition-decidable} that it is decidable whether,
for a formula $A \in \SForm{\vSigma}$,\, $\vGamma \LCon A$.
\qed
\end{proof}

\subsubsection*{Relational databases}

Having defined the notions of an relational language and a relational 
theory, we are ready to define the notion of a relational database in 
the setting of \pfoBDif.

A \emph{relational database} $\mathit{DB}$ is a triple 
$(R,\vTheta,\vXi)$, where:
\begin{itemize} 
\item
$R = (\vSigma,\SForm{\vSigma})$ is a relational language;
\item
$\vTheta$ is a relational theory for $R$;
\item
$\vXi$ is a finite subset of $\SForm{\vSigma}$.  
\end{itemize} 
$\vTheta$ is called the \emph{relational theory of $\mathit{DB}$} and
$\vXi$ is called the \emph{set of integrity constraints of 
$\mathit{DB}$}.

The set $\vXi$ of integrity constraints of a relational database 
$\mathit{DB} = (R,\vTheta,\vXi)$ can be seen as a set of assumptions about 
the relational theory of the relational database $\vTheta$.
If the relational theory agrees with these assumptions, then the
relational database is called consistent.

Let $R = (\vSigma,\SForm{\vSigma})$ be a relational language, and
let $\mathit{DB} = (R,\vTheta,\vXi)$ be a relational database.
Then $\mathit{DB}$ \emph{is consistent} iff, 
for each $A \in \SForm{\vSigma}$ such that $A$ is an atomic fact for $R$
or $A$ is of the form $\Not A'$ where $A'$ is an atomic fact for~$R$:
\[
\renewcommand{\arraystretch}{1.35}
\begin{tabular}{c} 
$\vTheta \scEnt A$ is provable only if $\vTheta,\vXi \scEnt \Cons A$ is 
provable.
\end{tabular}
\]

Notice that, if $\mathit{DB}$ is not consistent, $\vTheta,\vXi \scEnt A'$
is provable with the sequent calculus proof system of $\FOCL(\vSigma)$ 
for all $A'\in \SForm{\vSigma}$.
However, the sequent calculus proof system of $\pfoBDif(\vSigma)$ rules 
out such an explosion.

\subsubsection*{Models of relational theories}

The models of relational theories for a relational language 
$R = (\vSigma,\SForm{\vSigma})$ are structures of $\pfoBDif(\vSigma)$ of 
a special kind.

Let $R = (\vSigma,\SForm{\vSigma})$ be a relational language.
Then a \emph{relational structure for $R$} is a structure $\mathbf{A}$ 
of $\pfoBDif(\vSigma)$ such that:
\begin{itemize}
\item
$\Null\sp\mathbf{A} = \bot$;
\item
for all $d_1,d_2 \in \mathcal{U}\sp\mathbf{A}$: 
\begin{list}{}{\setlength{\leftmargin}{1em}}
\item[]
${=\sp\mathbf{A}}(d_1,d_2) \neq \VBoth$;
\item[]
if $d_1 \neq \bot$ and $d_1 \neq \bot$, then
${=\sp\mathbf{A}}(d_1,d_2) \neq \VNeither$;
\end{list}
\item
for all $d \in \mathcal{U}\sp\mathbf{A}$: 
\begin{list}{}{\setlength{\leftmargin}{1em}}
\item[]
if $d \neq \bot$, then
there exists a $c \in \Func{0}(\vSigma)$ such that 
${=\sp\mathbf{A}}(d,c\sp\mathbf{A}) = \VTrue$; 
\item[]
if $d = \bot$, then
there exists a $c \in \Func{0}(\vSigma)$ such that 
${=\sp\mathbf{A}}(d,c\sp\mathbf{A}) = \VNeither$;
\end{list}
\item
for all $c_1,c_2 \in \Func{0}(\vSigma)$: 
\begin{list}{}{\setlength{\leftmargin}{1em}}
\item[]
if
\begin{tabular}[t]{@{\,}l@{}}
${=\sp\mathbf{A}}({c_1}\sp\mathbf{A},{c_1}\sp\mathbf{A}) \neq \VNeither$ 
and 
${=\sp\mathbf{A}}({c_2}\sp\mathbf{A},{c_2}\sp\mathbf{A}) \neq \VNeither$
and
${=\sp\mathbf{A}}({c_1}\sp\mathbf{A},{c_2}\sp\mathbf{A}) = \VTrue$ 
\\ \hspace*{1em} or \\
${=\sp\mathbf{A}}({c_1}\sp\mathbf{A},{c_1}\sp\mathbf{A}) = \VNeither$ 
and 
${=\sp\mathbf{A}}({c_2}\sp\mathbf{A},{c_2}\sp\mathbf{A}) = \VNeither$,
\end{tabular}
\\
then $c_1 \equiv c_2$;
\end{list}
\item
for all $n \in \Nat$, for all $P \in \Pred{n+1}(\vSigma)$, for all
$d_1,\ldots,d_{n+1} \in \mathcal{U}\sp\mathbf{A}$:
\begin{list}{}{\setlength{\leftmargin}{1em}}
\item[]
$P^\mathbf{A}(d_1,\ldots,d_{n+1}) \neq \VNeither$.
\end{list} 
\end{itemize} 

The following corollary of the definitions of relational structure 
axioms for $R$ and relational structure for $R$ justifies the term 
``relational structure axioms''.
\begin{corollary}
Let $R = (\vSigma,\SForm{\vSigma})$ be a relational language, and
let $\mathbf{A}$ be a structure of $\pfoBDif(\vSigma)$.
Then $\mathbf{A}$ is a relational structure for $R$ iff,
for all assignments $\alpha$ in $\mathbf{A}$,
for all $A \in \mathit{RSA}(R)$,\,
$\Term{A}{\mathbf{A}}{\alpha} \in \DValue$.
\end{corollary}

Let $R = (\vSigma,\SForm{\vSigma})$ be a relational language, and
let $\vTheta$ be a relational theory for $R$.
Then all models of $\vTheta$ are relational structures for $R$ because 
$\mathit{RSA}(R) \subseteq \vTheta$.\,
$\vTheta$ does not have a unique model up to isomorphism.
$\vTheta$'s predicate completion axioms fail to enforce a unique model up 
to isomorphism.
However, identification of $\VTrue$ and $\VBoth$ in the models of 
$\vTheta$ yields uniqueness up to isomorphism. 

Let $R = (\vSigma,\SForm{\vSigma})$ be a relational language, and
let $\mathbf{A}$ be a relational structure for $R$.
Then we write $\nabla \mathbf{A}$ for the relational structure 
$\mathbf{A}'$ for $R$ such that:
\begin{itemize}
\item
$\mathcal{U}\sp{\mathbf{A}'} = \mathcal{U}\sp\mathbf{A}$;
\item
for each $c\in \Func{0}(\vSigma)$, 
$c\sp{\mathbf{A}'} = c\sp\mathbf{A}$; 
\item
for each $n \in \Nat$,
for each $P \in \Pred{n+1}(\vSigma)$, 
for each $d_1,\ldots,d_{n+1} \in \mathcal{U}\sp{\mathbf{A}'}$,
$P\sp{\mathbf{A}'}(d_1,\ldots,d_{n+1}) = 
 \left \{
 \begin{array}{l@{\;\;}l}
 \VTrue  & \mathrm{if}\; P\sp\mathbf{A}(d_1,\ldots,d_{n+1}) \in
                         \set{\VTrue,\VBoth} \\
 \VFalse & \mathrm{otherwise};
 \end{array}
 \right.$
\item
for each $d_1,d_2 \in \mathcal{U}\sp{\mathbf{A}'}$,
${=\sp{\mathbf{A}'}}(d_1,d_2) \,=\, {=\sp{\mathbf{A}}}(d_1,d_2)$.
\end{itemize}

\begin{theorem}
\label{theorem-unique-rmodel}
Let $R = (\vSigma,\SForm{\vSigma})$ be a relational language, 
let $\vTheta$ be a relational theory for $R$, and
let $\mathbf{A}$ and $\mathbf{A}'$ be models of $\vTheta$.
Then $\nabla \mathbf{A}$ and $\nabla \mathbf{A}'$ are isomorphic 
relational structures.
\end{theorem}
\begin{proof}
The proof goes in almost the same way as the proof of part~1 of 
Theorem~3.1 from~\cite{Rei84a}.
The only point of attention is that it may be the case that, for some 
$P \in \Pred{n+1}(\vSigma)$ and 
$c_1,\ldots,c_{n+1} \in \Func{0}(\vSigma)$ ($n \in \Nat$), 
either $\Term{P(c_1,\ldots,c_{n+1})}{\mathbf{A}}{\alpha} = \VTrue$ and
$\Term{P(c_1,\ldots,c_{n+1})}{\mathbf{A'}}{\alpha} = \VBoth$ or
$\Term{P(c_1,\ldots,c_{n+1})}{\mathbf{A}}{\alpha} = \VBoth$ and
$\Term{P(c_1,\ldots,c_{n+1})}{\mathbf{A'}}{\alpha} = \VTrue$.
But, if this is the case, 
$\Term{P(c_1,\ldots,c_{n+1})}{\mathbf{\nabla A}}{\alpha} = \VTrue$ and
$\Term{P(c_1,\ldots,c_{n+1})}{\mathbf{\nabla A'}}{\alpha} = \VTrue$.
\qed
\end{proof}

\begin{theorem}
\label{theorem-exists-rtheory}
Let $R = (\vSigma,\SForm{\vSigma})$ be a relational language, and
let $\mathbf{A}$ be a relational structure for $R$.
Then there exists a relational theory $\vTheta$ for $R$ such that 
$\mathbf{A}$ is a model of $\vTheta$.
\end{theorem}
\begin{proof}
The proof goes in the same way as the proof of part~2 of Theorem~3.1 
from~\cite{Rei84a}.
\qed
\end{proof}

\section{Query Answering Viewed through \pfoBDif}
\label{QUERY-ANSWERING}

In this section, queries applicable to a relational database and their
answers are considered from the perspective of \pfoBDif.
As a matter of fact, the queries introduced below are closely related to 
the relational-calculus-oriented queries originally introduced
by~\cite{Cod72a}.
Three kinds of answers to queries are introduced.
Two of them take the integrity constraint of the database into 
account.
They differ in their approach to deal with inconsistencies.
A discussion of these approaches can be found 
in~\cite[Section~8]{Mid22b}.

\subsubsection*{Queries}

As to be expected in the current setting, a query applicable to a 
relational database involves a formula of \pfoBDif.

Let $R = (\vSigma,\SForm{\vSigma})$ be a relational language.
Then a \emph{query for $R$} is an expression of the form
$(x_1,\ldots,x_n) \suchthat A$, where:
\begin{itemize} 
\item
$x_1,\ldots,x_n \in \SVar$;
\item
$A \in \SForm{\vSigma}$ and all variables that are free in $A$ are among
$x_1,\ldots,x_n$.
\end{itemize} 

Let $\mathit{DB} = (R,\vTheta,\vXi)$ be a relational database.
Then a query is \emph{applicable to $\mathit{DB}$} iff it is a query for
$R$.

\subsubsection*{Answers}

Answering a query with respect to a consistent relational database 
amounts to looking for closed instances of the formula concerned that 
are logical consequences of a relational theory.
The main issue concerning query answering is how to deal with 
inconsistent relational databases.

Let $R = (\vSigma,\SForm{\vSigma})$ be a relational language, 
let $\mathit{DB} = (R,\vTheta,\vXi)$ be a relational database, and
let $(x_1,\ldots,x_n) \suchthat A$ be a query that is applicable to 
$\mathit{DB}$.
Then an \emph{answer to $(x_1,\ldots,x_n) \suchthat A$ with respect to 
$\mathit{DB}$} is a $(c_1,\ldots,c_n) \in {\Func{0}(\vSigma)}^n$ for 
which 
$\vTheta \scEnt \subst{x_1 \assign c_1}\ldots\subst{x_n \assign c_n}A$
is provable.

The above definition of an answer to a query with respect to a database
does not take into account the integrity constraints of the database
concerned.

\subsubsection*{Consistent answers}

The definition of a consistent answer given below is based on the 
following:
\begin{itemize}
\item
the formula that corresponds to an answer is a logical consequence of 
some set of atomic facts and negations of atomic facts that are logical 
consequences of the relational theory of the database;
\item
in the case of a consistent answer there must be such a set that does 
not contain an atomic fact or negation of an atomic fact that causes the 
database to be inconsistent.
\end{itemize}

Let $R = (\vSigma,\SForm{\vSigma})$ be a relational language.
Then a \emph{semi-atomic fact for $R$} is an $A \in \SForm{\vSigma}$ 
such that $A$ is an atomic fact for $R$ or $A$ is of the form $\Not A'$ 
where $A'$ is an atomic fact for $R$. 

Let $R = (\vSigma,\SForm{\vSigma})$ be a relational language, 
let $\mathit{DB} = (R,\vTheta,\vXi)$ be a relational database, and
let $(x_1,\ldots,x_n) \suchthat A$ be a query that is applicable to 
$\mathit{DB}$. 
Then a \emph{consistent answer to $(x_1,\ldots,x_n) \suchthat A$ with 
respect to $\mathit{DB}$} is a
$(c_1,\ldots,c_n) \in {\Func{0}(\vSigma)}^n$ 
for which there exists a set
$\vPhi$ of semi-atomic facts for $R$ such that:
\begin{itemize}
\item
for all $A' \in \vPhi$,\, $\vTheta \scEnt A'$ is provable and
$\vTheta, \vXi \scEnt \Cons A'$ is provable; 
\item
$\vPhi, \mathit{RSA}(R) \scEnt
 \subst{x_1 \assign c_1}\ldots\subst{x_n \assign c_n}A$
is provable. 
\end{itemize}

The above definition of a consistent answer to a query with respect to 
a database is reminiscent of the definition of a consistent answer to a 
query with respect to a database given in~\cite{Bry97a}.
It simply accepts that a database is inconsistent and excludes the 
source or sources of the inconsistency from being used in consistent
query answering.

\subsubsection*{Strongly consistent answers}

The definition of a strongly consistent answer given below is not so
tolerant of inconsistency and makes use of consistent repairs of the 
database.
The idea is that an answer is strongly consistent if it is an answer 
with respect to every minimally repaired version of the original 
database.

Let $R = (\vSigma,\SForm{\vSigma})$ be a relational language, and
let $\vLambda \subseteq \SForm{\vSigma}$ be a finite set of atomic facts
for $R$.
Then, following~\cite{ABC99a}, the binary relation $\leq_\vLambda$ on the 
set of all finite sets of atomic facts for $R$ is defined by:
\[\vLambda' \leq_\vLambda \vLambda'' \mbox{ iff }
  (\vLambda \diff \vLambda') \union (\vLambda' \diff \vLambda) \subseteq
  (\vLambda \diff \vLambda'') \union (\vLambda'' \diff \vLambda)\;.
\]
Intuitively, $\vLambda' \leq_\vLambda \vLambda''$ indicates that the extent
to which $\vLambda'$ differs from $\vLambda$ is less than the extent to 
which $\vLambda''$ differs from $\vLambda$.

Let $R = (\vSigma,\SForm{\vSigma})$ be a relational language, 
let $\vLambda \subseteq \SForm{\vSigma}$ be a finite set of atomic facts
for $R$, and
let $\vXi$ is a finite subset of $\SForm{\vSigma}$. 
Then \emph{$\vLambda$ is consistent with $\vXi$} iff
for all semi-atomic facts $A$ for $R$,\,
$\mathit{RT}(R,\vLambda) \scEnt A$ is provable only if 
$\mathit{RT}(R,\vLambda),\vXi \scEnt \Cons A$ is provable.
We write $\mathit{Con}(\vXi)$ for the set of all finite sets of atomic 
facts for $R$ that are consistent with $\vXi$.

\sloppy
Let $R = (\vSigma,\SForm{\vSigma})$ be a relational language, 
let $\vLambda \subseteq \SForm{\vSigma}$ be a finite set of atomic facts
for $R$,
let $\mathit{DB} = (R,\mathit{RT}(R,\vLambda),\vXi)$ be a relational 
database, and
let $(x_1,\ldots,x_n) \suchthat A$ be a query that is applicable to 
$\mathit{DB}$. 
Then a \emph{strongly consistent answer to 
$(x_1,\ldots,x_n) \suchthat A$ with respect to $\mathit{DB}$} is a
$(c_1,\ldots,c_n) \in {\Func{0}(\vSigma)}^n$ such that, 
for each $\vLambda'$ that is $\leq_\vLambda$-minimal in 
$\mathit{Con}(\vXi)$,
$\mathit{RT}(R,\vLambda') \scEnt 
 \subst{x_1 \assign c_1}\ldots\subst{x_n \assign c_n}A$ is provable.
The elements of $\mathit{Con}(\vXi)$ that are $\leq_\vLambda$-minimal in 
$\mathit{Con}(\vXi)$ are called the \emph{repairs of $\vLambda$}.

The above definition of a strongly consistent answer to a query with 
respect to a database is essentially the same as the definition of a 
consistent answer to a query with respect to a database given 
in~\cite{ABC99a}.
It represents, presumably, the first view on what the repairs of an 
inconsistent database are. 
Other views on what the repairs of an inconsistent database are have 
been taken in e.g.~\cite{LB06a,GM11a,CFK12a,CLP18a,Ber19a}.

\subsubsection*{Null values in answers}

If $(c_1,\ldots,c_n)$ is an answer, consistent answer or strongly 
consistent answer to a query with respect to a database with null 
values, then each of $c_1$, \ldots, $c_n$ may be syntactically equal to 
$\Null$.
Such answers are called \emph{answers with null values}.
In most work on query answering with respect to a database with null 
values, answers with null values are not considered.
Notable exceptions are~\cite{IL84a,Lib16a}, where the view taken on null 
values in relational databases is different from the view taken in this 
paper.

\subsubsection*{Decidability}

The following theorem is a decidability result concerning being an 
answer to a query with respect to a database.
\begin{theorem}
\label{theorem-decidable-add-add}
\sloppy
Let $R = (\vSigma,\SForm{\vSigma})$ be a relational language, 
let $\mathit{DB} = (R,\vTheta,\vXi)$ be a relational database, and
let $(x_1,\ldots,x_n) \suchthat A$ be a query applicable to 
$\mathit{DB}$.
Then it is decidable whether, 
for $(c_1,\ldots,c_n) \in {\Func{0}(\vSigma)}^n$: 
\pagebreak[2]
\begin{itemize}
\item
$(c_1,\ldots,c_n)$ is an answer to 
$(x_1,\ldots,x_n) \suchthat A$ with respect to $\mathit{DB}$;
\item
$(c_1,\ldots,c_n)$ is a consistent answer to 
$(x_1,\ldots,x_n) \suchthat A$ with respect to~$\mathit{DB}$;
\item
$(c_1,\ldots,c_n)$ is a strongly consistent answer to 
$(x_1,\ldots,x_n) \suchthat A$ with respect to~$\mathit{DB}$.
\end{itemize}
\end{theorem}
\begin{proof}
Each of these decidability results follows immediately from 
Theorem~\ref{theorem-decidable-add} and the definition of the kind of 
answer concerned.
\qed
\end{proof}
As a corollary of Theorem~\ref{theorem-decidable-add-add}, we have that
the set of answers to a query, the set of consistent answers to a query, 
and the set of strongly consistent answers to a query are computable.

\subsubsection*{Example}

The example given below is kept extremely simple so that readers that 
are not initiated in the sequent calculus proof system of \pfoBDif\ can 
understand the remarks made about the provability of sequents.

Consider the relational database whose relational language, say $R$, has 
constant symbols $a$, $b$, $c$, $d$, and $\Null$ and ternary predicate 
symbol $P$, whose relational theory is the relational theory of which  
$P(a,b,\Null)$, $P(a,\Null,c)$, $P(a,\Null,d)$, and $P(b,c,d)$ are the 
atomic facts, 
and whose single integrity constraint is
$\CForall{x,y,z,y',z'}
  {(P(x,y,z) \CAnd P(x,y',z')) \SImpl y = y'}$.
Moreover, consider the query 
$x \suchthat
 \CExists{y,z,z'}{P(x,y,z) \CAnd P(x,y,z') \CAnd z \neq z'}$.
Clearly, the set of answers is $\set{a}$.

The sets of semi-atomic formulas that are logical consequences of the
relational theory and do not cause the database to be inconsistent 
include only one of $P(a,b,\Null)$, $P(a,\Null,c)$, and $P(a,\Null,d)$.
This means that, for each such set, say $\vPhi$, there is no constant 
symbol $k \in \set{a,b,c,d,\Null}$ such that 
$\vPhi,\mathit{RSA}(R) \scEnt \nolinebreak
 \subst{y \assign \nolinebreak k}
  (\CExists{x,z,z'}{P(x,y,z) \CAnd P(x,y,z') \CAnd z \neq z'})$ 
is provable. 
Hence, the set of consistent answers is $\emptyset$.

The repairs of $\set{P(a,b,\Null),P(a,\Null,c),P(a,\Null,d),P(b,c,d)}$ 
include only one of $P(a,b,\Null)$, $P(a,\Null,c)$, and $P(a,\Null,d)$.
This means that, for each repair, say $\vLambda$, there is no constant 
symbol $k \in \set{a,b,c,d,\Null}$ such that 
$\mathit{RT}(R,\vLambda) \scEnt \nolinebreak
 \subst{y \assign \nolinebreak k}
  (\CExists{x,z,z'}{P(x,y,z) \CAnd P(x,y,z') \CAnd z \neq z'})$ 
is provable. 
Hence, the set of strongly consistent answers is also $\emptyset$.

Now, consider the relational database that differs from the one 
described above only in that in the integrity constraint $y = y'$ is
replaced by $y \SEq y'$.
It is easy to see that in this case the set of answers and
the set of strongly consistent answers to the query described above 
remain the same, but the set of consistent answers becomes $\set{a}$.

Several examples in which null values play no role can be found 
in~\cite{Mid22b}.

\section{The Views on Null Values in Databases}
\label{REMARKS}

In this short section, some remarks are made on the different views 
that exist on null values in relational databases.

\subsubsection*{Our view}

The view taken in this paper on null values in relational databases can 
be summarized as follows:
\begin{itemize}
\item
in relational databases with null values, a single dummy value, called 
the null value, is used for values that are indeterminate;
\item
a value that is indeterminate is a value that is either unknown or 
nonexistent;
\item
independent of whether it is unknown or nonexistent, no meaningful 
answer can be given to the question whether the null value and whatever 
value, including the null value itself, are the same.
\end{itemize}
In the database literature, a null value for all values that are 
indeterminate, i.e.\ for all values that are either unknown or 
nonexistent, is usually called a \emph{no information} null value.
Moreover, the term \emph{inapplicable} is often used instead of 
nonexistent.

\subsubsection*{Other views}

Firstly, the view taken in this paper means that there are no separate 
null values for values that are unknown and values that are nonexistent.
Separate null values for values that are unknown and values that are 
nonexistent have for example been studied in~\cite{Vas79a,Ges90a}.
In those papers, query answering is based on a four-valued logic devised
to deal with the two different null values.
The logics concerned differ from each other and both differ a lot from 
\pfoBDif.

Secondly, the view taken in this paper means that the values for which 
the null value is a dummy value is not restricted to values that are 
unknown.
A single null value, for values that are unknown only, has for example 
been studied in~\cite{Cod79a,Bis83a,IL84a}.
In~\cite{IL84a} and later publications in which this kind of null value 
is considered, query answering with respect to a database with null 
values is usually approached in a way that is based on the idea that a 
database with null values represents a set of possible databases without 
null values.
This approach is reminiscent of the database repair approach to query 
answering with respect to an inconsistent database from~\cite{ABC99a}.

Thirdly, the view taken in this paper means that there are no multiple
null values.
Multiple null values, each for values that are unknown only, has for 
example been studied in~\cite{IL84a,Lib16a}.
In those papers, the same approach to query answering is applied as 
in~\cite{IL84a} to the case of a single null value.

A single null value for both values that are unknown and values that 
are nonexistent has also been studied before in~\cite{Zan84a,Kle03a}.
In~\cite{Zan84a}, query answering is based on a three-valued logic that 
is closely related to \pfoBDif.
In~\cite{Kle03a}, an attempt is made to apply the approach to query 
answering from~\cite{IL84a} to the case of a single null value for both 
values that are unknown and values that are nonexistent. 

The fact that so many views on null values in relational databases have 
been studied indicates that time and again the views already studied 
turned out to be unsatisfactory from one angle or another.
The view taken in this paper is primarily based on the consideration 
that it should be relatively simple from a logical perspective, taking 
into account that a database is anyhow no more than a representation of 
an approximation of a piece of reality.

\subsubsection*{The equality issue}

The view taken in this paper is essentially the same as the view taken 
in~\cite{Zan84a} and is based on similar considerations.
There is one aspect of this view that is often considered undesirable in 
work on null values in relational databases, namely the point that 
no meaningful answer can be given to the question whether the null value 
equals itself.
Made precise in the setting of this paper, it is the following that is 
often considered undesirable:
\[
\Term{t_1 = t_2}{\mathbf{A}}{\alpha} = \VNeither\quad \mathrm{if}\;\; 
\Term{t\sb1}{\mathbf{A}}{\alpha} = \Term{\Null}{\mathbf{A}}{\alpha}\; 
\mathrm{and}\;
\Term{t\sb2}{\mathbf{A}}{\alpha} = \Term{\Null}{\mathbf{A}}{\alpha}\;.
\]
The main reason that is given to consider this undesirable is that it 
may lead to counterintuitive answers to queries.
However, this is only so in the case where it is (implicitly) assumed 
that the null value is a dummy value for values that are unknown only.
In that case, the following is considered more appropriate:
\[
\Term{t_1 = t_2}{\mathbf{A}}{\alpha} = \VTrue\quad \mathrm{if}\;\; 
\Term{t\sb1}{\mathbf{A}}{\alpha} = \Term{\Null}{\mathbf{A}}{\alpha}\; 
\mathrm{and}\;
\Term{t\sb2}{\mathbf{A}}{\alpha} = \Term{\Null}{\mathbf{A}}{\alpha}
\phantom{\;.}
\]
(see e.g.~\cite{Lib16a}).

It is worth mentioning that the adapted equality predicate can be 
expressed in the setting of this paper.
Consider the abbreviation $t_1 \SEq t_2$ introduced earlier. 
As mentioned before, it follows from the definition of this abbreviation 
that:
\[
\begin{array}[t]{l@{\;\;}c@{\;\;}l}
\Term{t_1 \SEq t_2}{\mathbf{A}}{\alpha} & = &
 \left \{
 \begin{array}{l@{\;\;}l}
 \VTrue  & \mathrm{if}\; 
           \Term{t\sb1}{\mathbf{A}}{\alpha} = \bot\; \mathrm{and}\;
           \Term{t\sb2}{\mathbf{A}}{\alpha} = \bot \\
 \Meq\sp\mathbf{A}(\Term{t\sb1}{\mathbf{A}}{\alpha},
                   \Term{t\sb2}{\mathbf{A}}{\alpha})
         & \mathrm{otherwise}.
 \end{array}
 \right.
\end{array}
\]
Clearly, $\SEq$ corresponds to the adapted equality predicate.
This means that, although the definitions in Sections~\ref{DATABASE}
and~\ref{QUERY-ANSWERING} are based on the view that the null value is
a dummy value for both values that are unknown and values that are 
nonexistent, \pfoBDif\ allows for treating the null value as a dummy 
value for values that are unknown only.

\section{Concluding Remarks}
\label{CONCLUSIONS}

In this paper, a coherent logical view on relational databases and query 
answering is presented that takes into account the possibility that a 
database is a database with null values and the possibility that a 
database is an inconsistent database.
The presented view combines:
\begin{itemize}
\item
the proof-theoretic logical view of Reiter~\cite{Rei84a} on what is a 
relational database, a query applicable to a relational database, and an 
answer to a query with respect to a consistent relational database 
without null values;
\item
the view of Zaniolo~\cite{Zan84a} on null values in relational 
databases;
\item
the view of Bry~\cite{Bry97a} as well as the view of 
Arenas et al~\cite{ABC99a} on what is a consistent answer to a query 
with respect to an inconsistent relational database.
\end{itemize}
The view is expressed in the setting of the paraconsistent and 
paracomplete logic \pfoBDif.
In a logic that is not paracomplete, it would have been difficult to 
take properly into account the possibility that a database is a database 
with null values, and in a logic that is not paraconsistent, it would 
have been difficult to take properly into account the possibility that a 
database is an inconsistent database.

The main views on what is a consistent answer to a query with respect to 
an inconsistent relational database are the view of Bry~\cite{Bry97a} 
and the view of Arenas et al~\cite{ABC99a}.
The latter view is based on a notion of a minimal repair of an 
inconsistent database.
Most other views on consistent query answering, for example the views
presented in~\cite{LB06a,GM11a,CFK12a,CLP18a,Ber19a}, are based on a 
notion of a minimal repair of an inconsistent database that differs from 
the one from~\cite{ABC99a} by different choices concerning, among other 
things, the kinds of changes (deletions, additions, alterations) that 
may be made to the original database to obtain a repair and what is taken 
as the extent to which two databases differ.

There are four main views on null values in relational databases, namely
the view of Zaniolo~\cite{Zan84a}, two views of Imieli{\'n}ski and 
Lipski~\cite{IL84a}, and the view of Vassilion~\cite{Vas79a}.
Other views are mostly variations on one of these views.
The \emph{Codd-table} view from~\cite{IL84a}, in which there are no 
multiple null values, can be expressed in the setting of \pfoBDif\ 
because the single null value can be treated as in that view by using 
$\SEq$ instead of $=$.
In the \emph{V-table} view from~\cite{IL84a}, there may be multiple null
values (called marked nulls).
A variation of \pfoBDif\ is needed to express this view.
In the view from~\cite{Vas79a}, there are separate null values for 
values that are unknown and values that are nonexistent.
It is questionable whether \pfoBDif\ is suitable to express this view on 
null values in relational databases, in particular if the possibility 
that a databases is inconsistent has to be taken into account as well.

It is shown in this paper that, for each of the three kinds of answer 
introduced, being an answer to a query with respect to a relational 
database is decidable.
However, it is to be expected that query answering for these kinds of 
answers is intractable without severe restrictions on the integrity 
constraints and/or the use of null values.
A study of the computational complexity of query answering for these 
kinds of answers in the setting of \pfoBDif\ is an interesting option 
for future work.
 
\bibliographystyle{splncs03}
\bibliography{PCL}

\end{document}